\documentclass[preprints,editorial,accept,moreauthors,pdftex,10pt,a4paper]{Definitions/mdpi}

\pdfoutput=1

\firstpage{1}
\pubvolume{xx}
\issuenum{1}
\articlenumber{1}
\pubyear{2020}
\copyrightyear{2020}
\externaleditor{}
\history{}

\usepackage{mathptmx}
\usepackage{mathrsfs}
\usepackage{bbold}
\usepackage{amssymb}
\usepackage{bm}
\usepackage[LY1]{fontenc}
\newcommand\ket[1]{\,|#1\rangle}

\Title{Quantum Communication---Celebrating the Silver Jubilee of Teleportation}

\Author{Rotem Liss *\orcidA{} and Tal Mor}

\AuthorNames{Rotem Liss and Tal Mor}

\address[1]{Computer Science Department, Technion, Haifa 3200003, Israel; talmo@cs.technion.ac.il}

\corres{Correspondence: rotemliss@cs.technion.ac.il}

\keyword{quantum communication; quantum teleportation; quantum entanglement}

\begin{document}
\section{Introduction: Quantum Teleportation---Meaning and~Influence}
In 1993, Charles H. Bennett, Gilles Brassard, Claude Cr\'epeau,
Richard Jozsa, Asher Peres, and~William K. Wootters
published their seminal paper presenting quantum teleportation,
titled ``Teleporting an unknown quantum state via dual classical and
Einstein--Podolsky--Rosen channels''~\cite{teleport93}.
Their paper presents and answers the question
``Can we transmit an unknown quantum state
\emph{without physically sending it}?''
Namely, can we send enough information about our unknown quantum state,
in a way that would enable the receiver to obtain (i.e., regenerate) it?
Their paper provides a striking answer:
``Yes.\ An arbitrary state of a quantum bit (denoted by
$\ket{\psi} \triangleq \cos \left( \frac{\theta}{2} \right) \ket{0} +
e^{i\phi} \sin \left( \frac{\theta}{2} \right) \ket{1}$)
can be transmitted \emph{if} both the sender and the receiver
share a maximally entangled quantum state
(for example, the~\emph{singlet} state, denoted by
$\ket{\Psi^-} \triangleq \frac{\ket{01} - \ket{10}}{\sqrt{2}}$)
\emph{and} the sender can transmit classical messages
(only two standard/classical bits) to the receiver.''
This answer, which presented the \emph{quantum teleportation} protocol,
has revolutionized the field of quantum~communication.

Intuitively, the~teleportation paper proves the equivalence
``a quantum communication channel =
a shared entangled state + a classical communication channel''.
In particular,
``\emph{sending} an unknown state of \textbf{one}
 \emph{quantum} bit can be done by
\emph{sharing} (ahead of time) \textbf{one} maximally entangled state
of two \emph{quantum} bits +
\emph{sending} \textbf{two} \emph{classical} bits''.
The above equivalence is very important, because~quantum channels tend
to be much less reliable (and much more prone to losses and errors)
than classical channels; moreover, even if the sender and the receiver
only share (many) \emph{noisy} entangled states,
they can still employ quantum teleportation by first distilling
(a~fewer number of) nearly \emph{maximally}
entangled states~\cite{purify96}. (This method,
in particular, makes it possible to transmit arbitrarily faithful
quantum states over a \emph{noisy} quantum channel~\cite{purify96},
even without using quantum error-correcting codes~\cite{qec95}.)

\begin{figure}[H]
\centering
\includegraphics{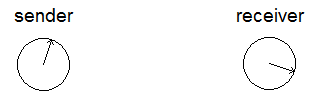}
\caption{We illustrate the power of quantum teleportation by 
representing the quantum bit as an \emph{arrow}
(a~two-dimensional arrow in this drawing;
a three-dimensional arrow in general) inside a unit sphere.
In the general three-dimensional case,
this representation is known as the \emph{Bloch sphere representation}.
The sender would like to transmit the arrow's direction to the receiver,
without physically sending the~arrow.}
\label{fig_arrows}
\end{figure}

To see how surprising this result is,
let us represent the quantum bit as an \emph{arrow}
directed at some arbitrary direction in the three-dimensional space
(see Figure~\ref{fig_arrows} for a two-dimensional illustration).
The~arrow's direction can be represented in spherical coordinates
by the two angles $\theta, \phi$
(note that $\theta, \phi$ are also the two angles that
appear in the mathematical representation
$\ket{\psi} \triangleq \cos \left( \frac{\theta}{2} \right) \ket{0} +
e^{i\phi} \sin \left( \frac{\theta}{2} \right) \ket{1}$
of the quantum bit).
Therefore, the~corresponding \emph{classical} question is
``Can we transmit the arrow's direction \emph{without physically
sending the arrow}?'' The obvious classical answer is
``Yes,~but~only if we send the real numbers $\theta, \phi$.''
Namely, in~the classical case, even when the sender \emph{knows}
the arrow's direction, a~very large number of classical bits must be
sent so that the receiver can reconstruct the approximate direction
of the arrow (the degree of precision dictates the number of sent bits;
infinite~precision requires an \emph{infinite} number of bits).
On the other hand, \emph{two} classical bits
would give us a very limited amount of information,
not allowing the receiver to recover the arrow's
direction at any reasonable amount of precision.
This is true even if the sender and the receiver share some information
in advance, assuming that the arrow's direction is chosen randomly and
independently of the shared~information. (In
a limited classical case, where the sender wants the receiver
to get the probability distribution of \emph{one} biased coin,
we can have some kind of ``classical teleportation'',
even if that distribution is unknown to the sender;
see details in~\cite{mor06}.)
\unskip\vspace{-8pt}

The \emph{quantum} case seems even worse:
if the sender holds the unknown quantum state $\ket{\psi}$
and wants to transmit it to the receiver, the~sender
apparently still has to send the two real numbers $\theta, \phi$.
Moreover, due to the peculiar properties of quantum mechanics,
those real numbers are now \emph{not even known} to the sender,
because the description of
$\ket{\psi} \triangleq \cos \left( \frac{\theta}{2} \right) \ket{0} +
e^{i\phi} \sin \left( \frac{\theta}{2} \right) \ket{1}$
is unknown to the sender.
(Note that the sender cannot discover the description of $\ket{\psi}$,
and any attempt to do so would irreversibly damage the quantum state.)
Nonetheless, the~quantum teleportation paper proves that by using
the extraordinary power of quantum entanglement, \emph{only two}
classical bits need to be~sent.

The teleportation paper is one of the most prominent examples of
the counterintuitive power of quantum communication; other
notable examples include quantum cryptography~\cite{bb84,bb84_journal},
violations of Bell's inequality~\cite{bell64}, and~even the basic
phenomena of quantum entanglement and EPR pairs~\cite{epr35}.

\section{The Discovery of Quantum Teleportation:
History, Notes, and~Stories}
Like any groundbreaking result, there are several interesting stories
surrounding the discovery of quantum teleportation.
Perhaps most interesting of all is the story of the actual invention
of quantum teleportation, as~recounted by Gilles Brassard and
printed here for the first time (except for an earlier personal account
in French~\cite{teleport_french05}):
\begin{quote}
``It~all started in August 1992, when I was attending the annual
CRYPTO conference. Charlie~Bennett gave me a paper that had appeared
in \emph{Physical Review Letters} one year earlier, saying~`I~think this will interest you'. Right he was! That was the
paper by Asher Peres and William (Bill) K.\ Wootters~\cite{PW91}
in which they considered the following problem: if~two participants
hold identical copies of an unknown quantum state~$\ket{\psi}$,
so that the state of their joint system is \mbox{$\ket{\psi}_\mathrm{A}
\otimes \ket{\psi}_\mathrm{B}$}, how much information can they discover
about~$\ket{\psi}$ if they are restricted to
local quantum operations and classical communication
(this is of course what became known later as~LOCC)? In~that paper,
Peres and Wootters studied so-called ping-pong protocols in which
more information can be obtained by increasing the number of interaction
rounds, but~they were unable to get quite as much information as
if the two identical quantum states were in the same location,
enabling the possibility of a joint measurement.
Their paper left open the following question: can LOCC measurements
provide as much information as joint measurements?

At~the time, I~had never met Peres or Wootters, and~in fact I~had
never heard of~them. I~met them both a few months later by~a pleasant
coincidence, at~the October 1992 \emph{Workshop on Physics and
Computation} held in Dallas. After~discussing the paper with its
authors, I~invited Bill to come to Montr\'eal to give a talk about it
the following month. Somehow, I~had a feeling this would be momentous,
and therefore I~invited Claude Cr\'epeau (who was in Paris at the time)
and Charlie Bennett to attend the talk at my expense. Richard
Jozsa was in the audience as well because he was my research assistant
at the time. After~Bill explained the conundrum, Charlie raised his
hand and asked an apparently inane question: `What~difference would
it make if the two participants shared an EPR pair?'\ (that's what we
called entanglement in those days). Not surprisingly, Bill replied
`I~don't know!' and then went on with his talk. Immediately afterwards,
we all moved to my office and brainstormed about Charlie's
question. By~the next morning, the~answer was clear: in the~presence of
entanglement, one~party teleports $\ket{\psi}$ to the other,
who then performs the optimal joint measurement. It~is fair to say
that we were able to invent quantum teleportation within less than
24~hours because none of us was trying to achieve this obviously
impossible~task! Of~course, we~realized that this invention was far
more important than the solution it offered to the problem at hand,
but I~don't think any of us anticipated how important it would
become. We~quickly invited Asher to join the collaboration and,
within eleven days, the~paper was submitted to \emph{Physical Review
Letters}. The~rest is history.''
\end{quote}

The writing process of that seminal paper was not exempt from dilemmas.
Gilles Brassard describes one of them---the length of the~paper:
\begin{quote}
``Once we had a version of the paper that we really liked, we noticed
that it was just a little too long for the then strict limit of
four pages imposed by \emph{Physical Review Letters}. We~could not
find anything that we would be comfortable leaving out.
That was when a devilish idea came to~me. Given that the type is
smaller in figure captions (8.5 points) than in the main text
(9.5 points), why not squeeze in some content there? We~relegated the
proof that successful teleportation of one qubit \emph{requires}
the transmission of two classical bits to what became a 27-line
caption for Figure~2 (see~\cite{teleport93}),
which saved exactly the required amount of
space to fit the paper snuggly in four pages. Ironically, we ended up
being the first paper of its issue, and~the space needed for the journal
header made us spill on a fifth page!''
\end{quote}

Another important dilemma was the order of the authors' names.
Readers unfamiliar with the advantages
and disadvantages of alphabetical order may not be able to understand
and appreciate the subtleties of the following story.
Alphabetical order for authors' names is customary in our
field, in~contrast to the ``contribution order'', which is conventional
in many others. There are researchers who participate only or
mostly in alphabetical-order papers, and~it is very important
to many of them to \emph{avoid combining the two methods}:
combining in that way could have a negative potential impact
on both their own research career
and the careers of their alphabetic~co-authors.

The original teleportation paper listed authors in alphabetical order.
Charles Bennett, who~frequently held the position of first author due to
his last name, with~many papers being cited as ``Bennett~et~al.'',
felt he was being over-credited in the eyes of
those accustomed to contribution order.
For this reason, at~some point before submission of the
teleportation paper, he suggested the use of reverse alphabetical order
for the authors, which would have placed Bill Wootters as first author.
This~idea was almost immediately rejected by Wootters himself.
Gilles Brassard, who has never once strayed from alphabetical order
throughout his entire career, told us years later that he felt so
strongly about this issue, that he would have withdrawn his name from
the paper had Bennett's reverse authorship idea been carried out.
Of course he could not have known this at the time,
but~taking himself off the paper might have prevented him from
sharing the Wolf Prize with Charles Bennett one quarter of a century
later. (Details about the Wolf Prize won by Bennett and Brassard,
which they received \emph{both} for quantum cryptography and quantum
teleportation, are provided at the end of the current section.)

Yet another dilemma was the \emph{name} of the new method.
Asher Peres objected to the original name ``teleportation''
because it mixed the Greek prefix ``tele-''
with the Latin-based root ``port''.
Peres~suggested the alternative name ``telepheresis'',
but the other authors disagreed, so the name remained ``teleportation''.

The quantum teleportation paper received excellent
reviews before being accepted to \emph{Physical Review Letters}.
One of the reviewers, N. David Mermin, described the paper as
a ``charming, readable, thought-provoking paper'', and~predicted that
``this novel method [\ldots] will become an important one of
the intellectual tools available to anybody [\ldots]''
(see Mermin's paper~\cite{mermin04}, where he disclosed
his full referee report on the quantum teleportation paper).
Finally, the~paper was published
on 29 March 1993~\cite{teleport93}
in \emph{Physical Review Letters}, profoundly advancing the field
of quantum communication and bringing new researchers
to the fast-evolving field of quantum information processing.
In particular, it~influenced Tal Mor, who is one of the authors
of this editorial, as~he describes~below.
\begin{quote}
When the teleportation paper was published (1993),
I was an M.Sc.~student in Tel Aviv University (Israel)
in the group led by Yakir Aharonov, together with Sandu Popescu
(who~was a Ph.D.~student at the time) and Lev Vaidman
(who was a postdoctoral researcher). All three of us (Popescu,
Vaidman, and~I) were extremely excited about the teleportation paper:
Popescu suggested a method~\cite{popescu95}
leading to the first experimental realization
of quantum teleportation~\cite{teleport_exp97a} (note that
quantum teleportation was experimentally demonstrated
in 1997--1998 by three research
groups~\cite{teleport_exp97a,teleport_exp97b,teleport_exp98});
Vaidman suggested teleportation of continuous quantum
variables~\cite{vaidman94}, leading to a theoretical
extension~\cite{teleport_BK98} and its experimental
realization~\cite{teleport_exp98};
and I decided to start my Ph.D.~with
Asher Peres, who was one of the teleportation paper's authors.
Although I concentrated on quantum cryptography,
I also gave a lot of thought to quantum teleportation:
I presented teleportation as a special case of POVM
(generalized measurements) in my first talk at an international
conference~\cite{telepovm96_arxiv,telepovm96,telepovm04},
and I suggested how a classical variant of teleportation
could look like (a~concept I published years later~\cite{mor06}).

The quantum teleportation paper and its experimental realizations
intrigued not only scientists, but~also media reporters. When one of
them asked Peres whether quantum teleportation teleports only a person's
body, or~also the soul, Peres answered that it teleports \emph{only} the
soul~\cite{peres05}---a funny, thought-provoking reply from
someone like Peres, who enjoyed describing himself as a devout atheist!

During my Ph.D. and postdoctoral research, I became acquainted with
all six authors of the teleportation paper. I even asked them to
autograph an original reprint of the paper---so~I now own the only
copy of the quantum teleportation paper signed by all six
co-authors! (Admittedly, it was pretty hard to obtain
this signed copy. Unfortunately, Gilles Brassard, who was the last
co-author to sign, lost the copy signed by
the five other co-authors in his office; later, he sent me
an e-mail including the ``good news''---that he found the signed copy of
the teleportation paper---and the ``bad news''---that he lost it again;
finally, he found it \emph{again}, signed it,
and immediately mailed it from Canada to me in Israel,
and I received it. Then,~\emph{I}~lost it in \emph{my} office\ldots{}
where I may find it again some day.)

Subsequently, I had two opportunities to celebrate the quantum
teleportation paper and honor some of its authors at my institution
(Technion, Haifa, Israel): when I organized the QUBIT~2003 conference,
celebrating 10 years of quantum teleportation,
with Asher Peres as the guest of honor~\cite{qubit03};
and when I organized the QUBIT~2018 conference,
celebrating the Wolf Prize of Charles Bennett
and Gilles Brassard, with~both of them
as the keynote speakers~\cite{qubit18}.
\end{quote}

Charles Bennett and Gilles Brassard won
the 2018 Wolf Prize in physics ``for founding and advancing
the fields of Quantum Cryptography and Quantum Teleportation''.
The jury panel acknowledged
the enormous importance of quantum~teleportation:
\begin{quote}
``In the 1990's they [Bennett and Brassard],
together with four colleagues, invented quantum teleportation
which allows the communication of quantum information over classical
channels, also a task previously believed to be impossible.
Two decades after their proposal, quantum teleportation has
now been demonstrated over distances exceeding 1000 kilometers
and is clearly destined to play a major role
in future secure communications.''~\cite{wolf18}
\end{quote}

\section{The Papers in This Special~Issue}
This special issue is dedicated to celebrating the silver jubilee of
the seminal teleportation paper, and~it features contributions from
various areas of quantum~communication.

Francesco De Martini and Fabio Sciarrino, in~their paper ``Twenty years
of quantum state teleportation at the Sapienza University in
Rome''~\cite{si1_DMC}, review various experiments of quantum
teleportation that were conducted at the Sapienza University in Rome,
ranging from the \emph{first} teleportation experiment (1997) to several
variations and generalizations of teleportation, such as
active teleportation and quantum machines based on~teleportation.

Nicolas Gisin, in~his paper ``Entanglement 25 years after quantum
teleportation: Testing joint measurements in quantum
networks''~\cite{si2_gisin}, discusses quantum entanglement from
an unusual perspective: that of entangled \emph{measurements}
rather than entangled \emph{states}. In~particular, Gisin raises
the question of whether entangled measurements can be used for
generating non-classical output correlations in various quantum
networks, and~suggests a few candidates
that may present such non-classical~correlations.

Gilles Brassard, Luc Devroye, and~Claude Gravel, in~their paper
``Remote sampling with applications to general entanglement
simulation''~\cite{si3_BDG}, provide a (classical) sampling scheme:
their~scheme allows the user to sample
\emph{exactly} from a discrete probability distribution
when the defining parameters of this probability distribution
are partitioned between several remote parties.
Furthermore, they apply their sampling scheme to the classical
simulation of quantum entanglement measurements
in the most general scenario,
and analyze its communication~complexity.

William K. Wootters, in~his paper ``A classical interpretation of
the Scrooge distribution''~\cite{si4_wootters},
shows how to derive a special \emph{quantum} ensemble of pure states,
known as the ``Scrooge ensemble'' (or ``Scrooge distribution''),
from a \emph{classical} communication scenario.
Specifically, he proves that a real-amplitude variant
of the Scrooge distribution naturally appears in
a classical communication scheme, and~that the standard
(complex-amplitude) Scrooge distribution appears in
a modified version of the same communication~scheme.

Michel Boyer, Rotem Liss, and~Tal Mor, in~their paper
``Attacks against a simplified experimentally feasible
semiquantum key distribution protocol''~\cite{si5_BLM},
explore the security of a semiquantum key distribution (SQKD)
protocol that seems easy to implement in practice. In~particular,
they analyze a simplified variant of the previously
published ``Mirror'' SQKD protocol, and~prove that unlike
the original Mirror protocol (which was proved completely robust),
its simplified variant is completely insecure
if the tolerated loss rate is~high.

Kan Wang, Xu-Tao Yu, Xiao-Fei Cai, and~Zai-Chen Zhang, in~their paper
``Probabilistic~teleportation of arbitrary two-qubit quantum state
via non-symmetric quantum channel''~\cite{si6_WYCZ},
propose a variant of quantum teleportation:
their scheme allows teleporting an arbitrary two-qubit state
from Alice to Bob, given that Alice and Bob share one
partially entangled pure three-qubit state and one partially entangled
pure two-qubit state. Their teleportation scheme is probabilistic
and unambiguous: namely, it may fail with constant probability,
but the users know whether it succeeded or~failed.

We hope that the papers in this special issue give insight regarding
the different areas of quantum communication---most notably, quantum teleportation, quantum entanglement,
and quantum~cryptography.

\vspace{6pt}

\funding{The work of T.M. and R.L. was supported in part
by the Israeli~MOD.}

\acknowledgments{We thank all the authors who submitted their
contributions to this special issue. We~acknowledge all the anonymous
reviewers and the editorial staff of Entropy for their ongoing
support. We~also thank Gilles Brassard for his invaluable help
regarding all parts of this editorial and for writing
his historical account on the invention of quantum~teleportation.}

\conflictsofinterest{The authors declare no conflict of~interest.}

\reftitle{References}
\externalbibliography{yes}
\bibliography{editorial_arxiv}

\end{document}